\documentclass[journal = apchd5, manuscript = article, layout = traditional]{achemso}

\usepackage{graphicx}
\usepackage{amssymb}
\usepackage{amsmath}
\usepackage{color}

\author{Nojoon Myoung}
\affiliation{Department of Physics Education, Chosun University, Gwangju 61452, Republic of Korea}
\alsoaffiliation{Center for Theoretical Physics of Complex Systems, Institute for Basic Science, Daejeon 34051, Republic of Korea}
\author{Hee Chul Park}
\affiliation{Center for Theoretical Physics of Complex Systems, Institute for Basic Science, Daejeon 34051, Republic of Korea}
\author{Ajith Ramachandran}
\affiliation{Center for Theoretical Physics of Complex Systems, Institute for Basic Science, Daejeon 34051, Republic of Korea}
\author{Elefterios Lidorikis}
\affiliation{Department of Materials Science and Engineering, University of Ioannina, Ioannina 45110, Greece}
\author{Jung-Wan Ryu}
\affiliation{Center for Theoretical Physics of Complex Systems, Institute for Basic Science, Daejeon 34051, Republic of Korea}
\email{jungwanryu@gmail.com}	

\title{Flat-band localization and self-collimation of light in photonic crystals}


\date{\today}

\begin{document}

\begin{tocentry}
\includegraphics[width=6.0cm]{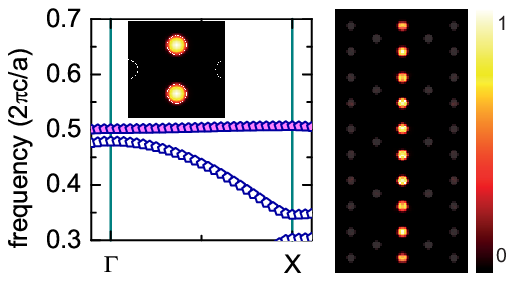}
\end{tocentry}

\begin{abstract}
We investigate the optical properties of a photonic crystal composed of a quasi-one-dimensional flat-band lattice array through finite-difference time-domain simulations. The photonic bands contain flat bands (FBs) at specific frequencies, which correspond to compact localized states as a consequence of destructive interference. The FBs are shown to be nondispersive along the $\Gamma\rightarrow X$ line, but dispersive along the $\Gamma\rightarrow Y$ line. The FB localization of light in a single direction only results in a self-collimation of light propagation throughout the photonic crystal at the FB frequency.
\end{abstract}

Photonic crystals (PCs), artificial materials governed by periodic dielectric functions, have been thoroughly studied for applications covering diverse spheres of interest. One key feature of PCs is the ability to manipulate light propagation through them---properties of electromagnetic waves inside PCs can be influenced by designing a periodic distribution of the dielectric function. Similar to electronic bands in solids, photonic bands exhibit band gaps where photonic states are prohibited, resulting in the prohibition of light propagation in a certain frequency range. Owing to the existence of photonic band gaps, PCs have been widely exploited as waveguides,\cite{Mekis1996,Foresi1997,Broeng1999,Hood2016,Faggiani2016,Jandieri2017} solar cells,\cite{Zeng2006,Bermel2007,Florescu2008,Mutitu2008,Colodero2009,Liu2016} lasesr\cite{Painter1999,Meier1999,Park2004,Hirose2014,Wu2015,Chen2016,Yokoo2017,Hwang2017} etc. In particular, one intriguing aspect of PCs is a localization of light, where a primary approach is to introduce a defect or imperfection to a PC that leads to localized modes at the defect, regarded as analogous to microcavity modes.\cite{Villeneuve1996,Painter1999,Bayindir2000,Regensburger2013,Faggiani2016,Gao2016}

Light localization in PCs has been an interesting topic in terms of both fundamentals and related applications. In 2D PCs composed of optical cavities, for instance, Anderson localization of light can be experimentally confirmed by introducing a random disorder to the PC;\cite{Schwartz2007,Topolancik2007,Liu2014,Faggiani2016,Roque2017,Vasco2017} notably, the stronger the localization, the less diffractive the light propagation through the disordered PC.\cite{Schwartz2007,Levi2011} In order to obtain stronger light localization in disordered PCs, a larger level of disorder is required.\cite{Schwartz2007,Hsieh2015} Observation of such strong light localization, moreover, requires confirmation through statistical means, i.e., ensemble average over many disordered PCs. Thus, it is not feasible to make use of Anderson localization in device applications, despite its non-diffusive features in terms of photon transport.

Recently, a novel type of localization has begun to attract attention, the so-called \textit{compact localized states} (CLSs) proposed by lattice models.\cite{Flach2014,Molina2015,Khomeriki2016,Leykam2017,Maimaiti2017} In specific quasi-1D lattices, such as cross-stitch, tunable diamond, etc., dispersion relations exhibit flat bands (FBs) over the whole Brillouin zone, corresponding to CLSs, or in other words a delta-function-like localization in real space.\cite{Flach2014,Molina2015,Baboux2016,Khomeriki2016,Leykam2017,Maimaiti2017,Travkin2017,RojasRojas2017,Perchikov2017} Similarly, FBs and CLSs are found in specific 2D lattices like Lieb and Kagome, among others.\cite{Mukherjee2015,Xia2016,Leykam2017,RojasRojas2017} The existence of CLSs has been experimentally demonstrated in PCs,\cite{Mukherjee2015,Xia2016,Travkin2017,RojasRojas2017} polaritonic systems,\cite{Baboux2016} and mechanical lattices.\cite{Perchikov2017} Contrary to Anderson localization, there is no need for statistical tasks to demonstrate CLSs because they require no disorder. Indeed, it has been experimentally shown that flat-band modes are non-diffractive, as diffraction is prohibited via destructive interference.\cite{Mukherjee2015,Weimann2016,Travkin2017} However, since flat-band modes are a consequence of destructive interference between amplitudes residing on specific sites in PCs, the corresponding CLSs are expected to be easily perturbed by inhomogeneity, and thus require sophisticated fabrication and measurement techniques.

In this paper, we investigate the photonic band structure and optical properties of a 2D PC, in the form of, an array of quasi-1D flat-band lattices, using finite-difference time-domain (FDTD) simulations. We demonstrate that localization of light corresponding to the 1D flat-band mode can remain compact even in the presence of coupling between individual quasi-1D lattices. The CLSs are found to be odd symmetric with respect to the transverse direction of the lattice. The photonic band structure of the 2D PC exhibits an anisotropic nature, following anisotropic couplings between lattice sites. Notably, for a specific frequency corresponding to the CLSs, we reveal a self-collimation of light propagation with a given out-of-phase excitation, thanks to strong localization in one direction only. Understanding defect-free localization of light in PCs paves the way for enhancing performance of optical devices, telecommunications, and sensing applications.

\section{Results and discussion}

\begin{figure}
\includegraphics[width=8.5cm]{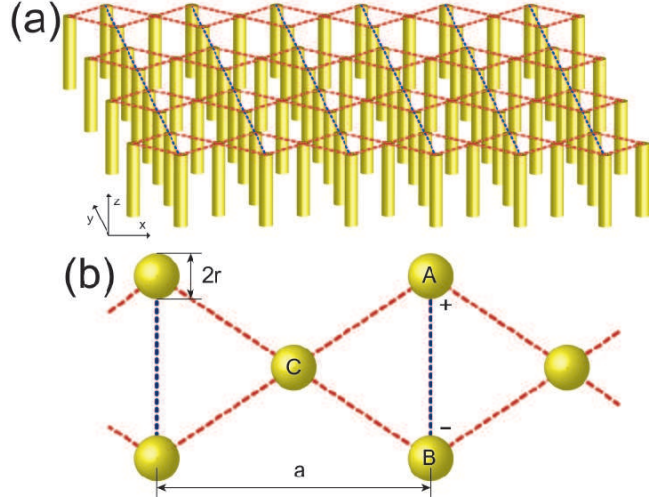}
\caption{(a) Schematic view of the photonic crystal considered. Dielectric nanopillars are arranged in two dimensions forming an array of quasi-one-dimensional ``tunable diamond lattice'' denoted with red dashed lines. There are two kinds of interactions in the photonic crystal: red dashed lines represent interactions between different sublattices, and blue dotted lines indicate interactions between the same sublattices. (b) Lattice model for the tunable diamond lattice. In a unit cell, electric fields guided in the majority sublattices (A and B) result in destructive interference at the minority sublattice (C), corresponding to compact localized states (CLSs) at A and B.} \label{fg:model}
\end{figure}

\subsection{Photonic bands and transmittance spectra}

Here, we consider a 2D PC structure where infinitely long dielectric nanopillars are arranged periodically to form a specific lattice, as displayed in Fig. \ref{fg:model}(a). Such a lattice structure is described by an array of quasi-1D tunable diamond (TD) lattices. Let us notice that the dielectric nanopillars in the array are infinitely long in the vertical direction, so that we may regard it as a 2D PC, assuming homogeneous electromagnetic fields along the $z$ direction.

The unit cell of the 2D PC contains three nanopillars, which can be split into two sublattices: majority (A and B) and minority (C), as depicted in Fig. \ref{fg:model}(b). Note that coupling between A (or B) and C differs from that between A and B. All dielectric nanopillars are identical with radius $r$, and the period of the TD lattice is $a$. In our 2D PC, this quasi-1D TD lattice is repeated with the period $a$, forming a square unit cell with three sublattices.

\begin{figure*}
\includegraphics[width=16cm]{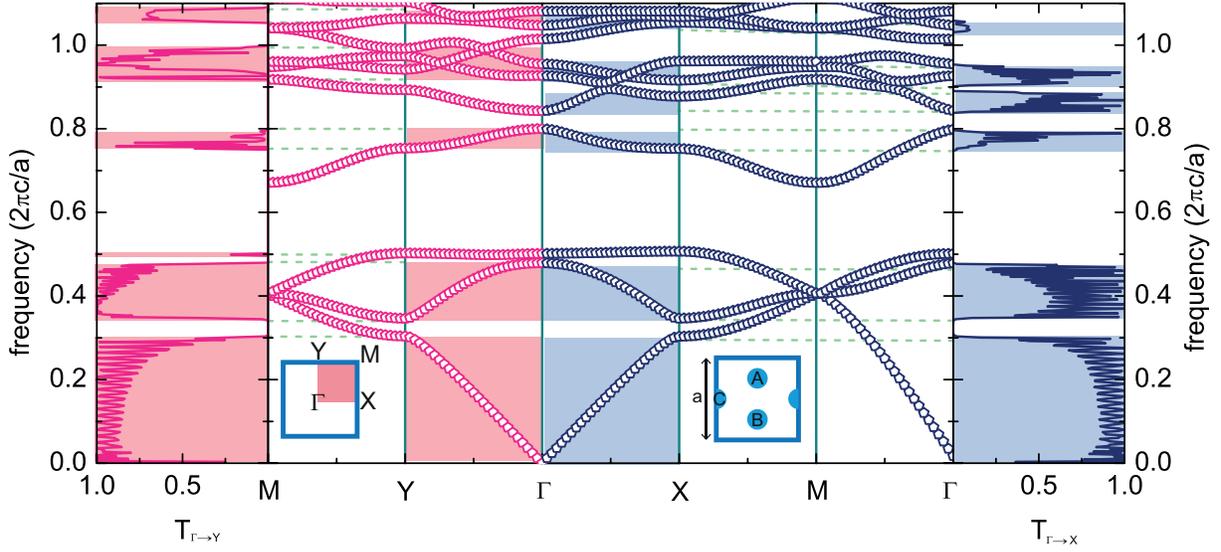}
\caption{ Photonic band structure of the 2D PC for TE polarization, with $r=0.1~a$ and $\epsilon=12.9$. Insets: (bottom left) The paths in the BZ followed in the band structure plot, and (bottom right) unit cell of the 2D PC used in calculations. Leftmost (rightmost) pannel: Transmittance spectra for the 2D PC with a homogeneous plane wave along the $y$ ($x$) direction. Shaded regions represent frequency ranges allowing for transmission of light. Dashed lines are eye guides to represent transmission gap edges.} \label{fg:band}
\end{figure*}

\begin{figure*}
\includegraphics[width=16cm]{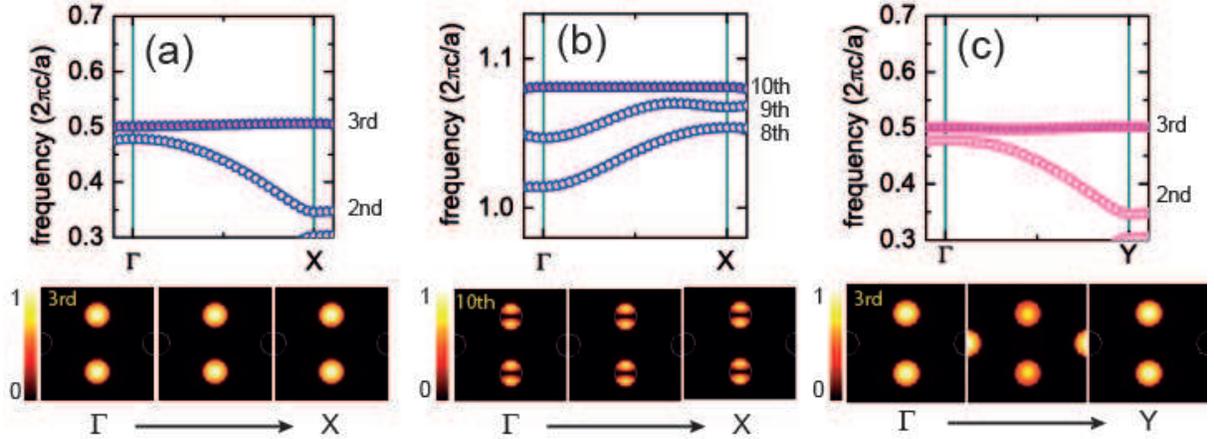}
\caption{ Enlargements of the photonic bands around (a) the third band along the $\Gamma\rightarrow X$ line, (b) the tenth band along the $\Gamma\rightarrow X$ line, and (c) the third band along the $\Gamma\rightarrow Y$ line, respectively. Lower pannels: Field intensity evolutions $\left|E\right|^{2}$ of the highlighed bands along the given symmetric point lines: (a) and (b) $\Gamma\rightarrow X$, and (c) $\Gamma\rightarrow Y$. Bright and dark colors indicate strong and weak field intensity, respectively.} \label{fg:FBdist}
\end{figure*}

In order to determine the presence of the FBs, we investigate the optical properties of the 2D PC by numerical simulations in both frequency and time domains, using freely available MPB and MEEP software packages, respectively\cite{Johnson2001,Oskooi2010}. Figure \ref{fg:band}(a) shows the photonic band structure for TE polarization ($E_{x}=E_{y}=0$ but $E_{z}\neq0$) of the PC calculated in the 2D Brillouin zone. The PC exhibits a wide band gap from $f\simeq0.5$ to 0.67, a narrow band gap from $f\simeq0.8$ to 0.84, and so on. Here, we focus on the photonic bands along the $\Gamma\rightarrow X$ line in momentum space, which corresponds to the one-dimensional nature of the quasi-1D TD lattice laid along the $x$ direction in real space. Indeed, one can see that there are two FBs at specific frequencies $f\simeq0.5$ and 1.08, as shown in Fig. \ref{fg:FBdist}(a) and (b). These FBs, which correspond to the third and tenth bands, are found to be flat only along the $\Gamma\rightarrow X$ line. Indeed, we achieve strongly localized electromagnetic waves in A and B, resulting in zero field intensity in C. (see lower pannels in Fig. \ref{fg:FBdist}(a) and (b))  

Since the FB reflects the existence of CLSs, propagation of light is expected to be prohibited at the flat-band frequencies in PCs. In order to examine the role of FBs for light propagation through the PC, we investigate the optical responses of the PC to a plane wave with TE polarization. We now consider a finite length of the PC in the $x$ direction with a periodic boundary condition in the $y$ direction. Twenty-one unit cells along the $x$ direction are taken into account in the FDTD simulation, a number believed to be large enough to reflect the photonic band structure that we found in the infinite PC. We put a plane wave source sufficiently far away from one facet of the finite PC, and calculate the energy flux of the transmitted wave throughout the PC to yield optical transmittance as a ratio of energy fluxes between incident and transmitted waves. The rightmost and leftmost pannels of Fig. \ref{fg:band} show the transmittance spectra $T_{\Gamma\rightarrow X}$ and $T_{\Gamma\rightarrow Y}$ as functions of the incident plane wave frequency in the $x$ and $y$ directions, respectively. $T_{\Gamma\rightarrow X}$ exhibit photonic band gaps in the specific frequency ranges where transmittance drops to zero. Overall, the photonic band gaps are well reflected in the transmittance spectra. However, we observe that certain photonic bands along the $\Gamma\rightarrow X$ line seem to have no contribution to $T_{\Gamma\rightarrow X}$ . As displayed by the shaded regions in Fig. \ref{fg:band}, the third, seventh, eighth, and tenth bands seem to not be involved in $T_{\Gamma\rightarrow X}$. The reason for these inactive photonic bands will be discussed in the next subsection.

On the other hand, let us consider optical transmittance $T_{\Gamma\rightarrow Y}$ through the 2D PC with a plane wave source propagating along the $y$ direction. Because the unit cell considered in this study contains three sublattices that are anisotropic (i.e., the unit cell has no symmetry with respect to 90$^{\circ}$ rotation), the PC exhibits an anisotropic photonic band structure (Fig. \ref{fg:band}) where the symmetric point $Y$ replaces $X$ in Fig. \ref{fg:band}(a). Overall, $T_{\Gamma\rightarrow Y}$ is similar to $T_{\Gamma\rightarrow X}$, but a distinguishing feature is that the third band along $\Gamma\rightarrow Y$ now contributes to optical transmittance. As highlighted in Fig. \ref{fg:FBdist}(c), the third band still seems to be flat over $\Gamma\rightarrow Y$. However, by investigating the related field intensities along the $\Gamma\rightarrow Y$ line, it is revealed to be a dispersive band, as displayed in the lower panel of Fig. \ref{fg:FBdist}(c), so that the electromagnetic wave can be transferred from A and B to C, and vice versa.

\subsection{Field distributions and compact localized states}

The optical property analysis results of the 2D PC in this study introduce two important questions: i) how the flat bands differ from the dispersive bands, and ii) why certain photonic bands, including the FBs, have no contribution to the optical transmittance. To answer these questions, we analyze field distribution in the unit cell of the PC for the resulting photonic bands.

\begin{figure}
\includegraphics[width=8.5cm]{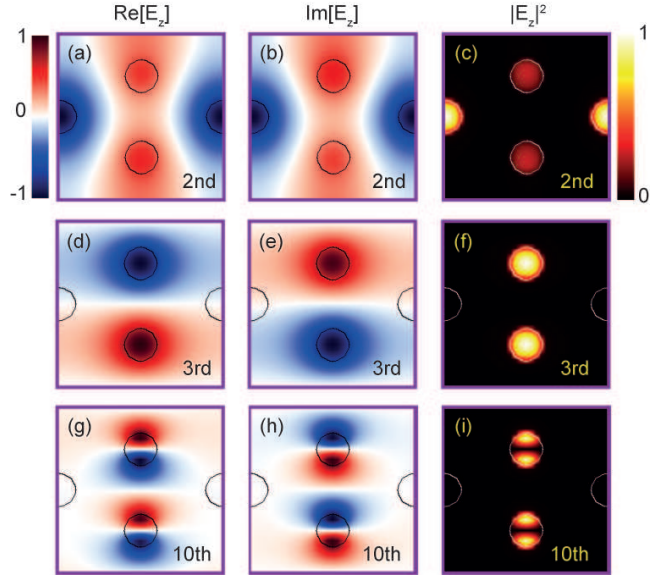}
\caption{(a), (b), and (c) Electric fields and field intensity distribution at $\Gamma$ in a unit cell of the 2D PC for the 2nd band. (d), (e), and (f) Electric fields and field intensity distribution at $\Gamma$ for the 3rd band. (g), (h), and (i) Electric fields and field intensity distribution for the 10th band. Red and blue colors indicate positive and negative electric fields, and bright and dark colors indicate strong and weak field intensity, respectively.} \label{fg:fieldFB}
\end{figure}

We first compare the electric field distributions of the flat and dispersive bands. As shown in Fig. \ref{fg:fieldFB},  both dispersive and FBs exhibit a distinguishing distributions of field intensity $\left|E\right|^{2}$. The second band, which is found to be dispersive in Fig. \ref{fg:band}(a), indeed has a state that spreads over the unit cell, whereas the third band supports CLSs as $\left|E\right|^{2}$ is strongly localized in A and B nanopillars only. Such a localization of the electric field results from destructive interference in the C nanopillar as the electric fields in A and B have the opposite sign. On the other hand, the dispersive band comprises the same sign electric fields in A and B, causing constructive interference in C. It is worth mentioning that the dispersive and FBs can be characterized by even and odd parities in the $y$ direction of the unit cell, respectively.

As aforementioned, two FBs were found in the given frequency range $f=0$ to $1.1$, corresponding to the third and tenth bands. Like the third band, the tenth band also exhibits a strong localization of electric field intensity. By comparing the electric fields in Fig. \ref{fg:fieldFB}(d) and (e), and (g) and (h), one can easily see that the tenth FB is attributed to odd parity with respect to the $y$ direction like the third band. Further, the tenth band occurs as a consequence of destructive interference between excited modes in A and B nanopillars, contrary the to fundamental modes for the third band.

\begin{figure*}
\includegraphics[width=16.0cm]{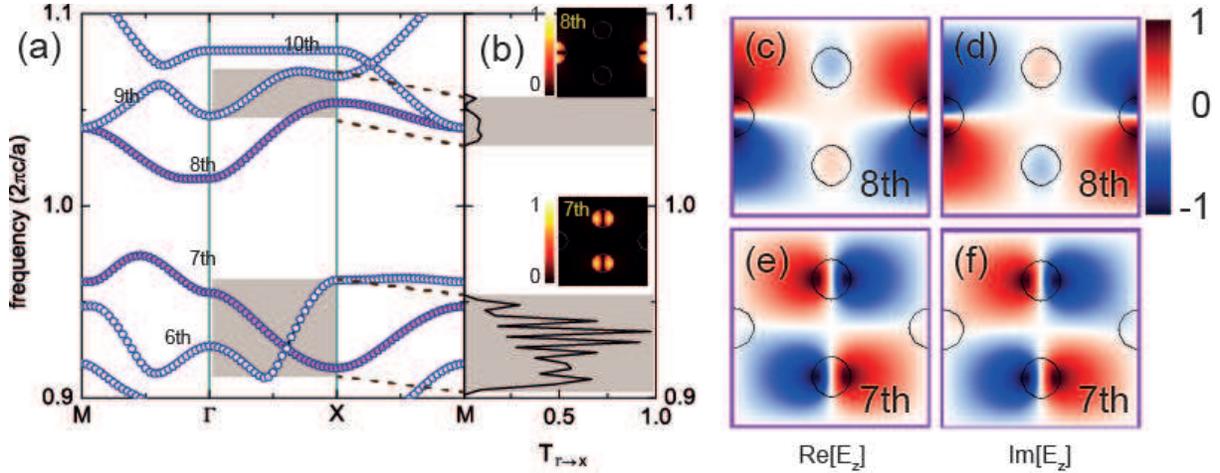}
\caption{(a) Photonic band structure of the 2D PC in the frequency range $f=0.9$--1.1 for TE polarization. (b) Transmittance spectra for the PC with a homogeneous plane wave along the $x$ direction. Shaded regions represent frequencies with finite transmittance. Insets: Field intensities at $\Gamma$ for the 7th (bottom) and 8th (top) bands, respectively. Bright and dark colors indicate strong and weak intensities. (c) and (d) Electric field distributions at $\Gamma$ for the 8th band. (e) and (f) Electric field distributions at $\Gamma$ for the 7th band. Red and blue correspond to positive and negative values of the electric fields.} \label{fg:fieldodd}
\end{figure*}

At this time, it should be noted that our electromagnetic wave source is provided as a plane wave propagating along the $x$ direction, in other words, the plane wave source is characterized as even parity with respect to the $y$ direction. This even-parity plane wave source does not excite the FBs, so that the FBs (the third and tenth bands) are not reflected in the transmittance spectra, as shown in Fig. \ref{fg:band}. This is because the FBs are dark modes with respect to the plane-wave source. Similarly, the seventh and eighth bands are also dark, as highlighted by color-filed symbols in Fig. \ref{fg:fieldodd}(a). By investigating their electric field patterns, we reveal that both the seventh and eighth bands indeed have odd parity with respect to the $y$ direction (see Fig. \ref{fg:fieldodd}). Here, let us notice that there are slight mismatches between the shaded regions in Fig. \ref{fg:fieldodd}(a) and (b), which are because of finite-resolution errors.

\subsection{Self-collimation of light propagation}

Due to the anisotropy of the 2D PC, the transmittance spectra with an incident wave along the $y$ direction are different from those for the $x$ direction (see Fig. \ref{fg:band}). In particular, it is worthwhile to note that $T_{\Gamma\rightarrow Y}$ is nonzero at frequencies $f\simeq0.5$ and 1.08, corresponding to the FBs along the $\Gamma\rightarrow X$ line. In other words, we expect anisotropic light propagation throughout the 2D PC at the flat-band frequencies.

Now, we consider  point sources inside the 2D PC instead of a plane wave source outside the PC. The point source radiates isotropic light with a frequency precisely set at $f=0.501107$, which is the frequency of the fundamental FB (third band). In order to examine the effects of parity, we place two monochromatic point sources at the centers of A and B nanopillars by changing their phase difference. Here, we stimulate a finite size (21$\times$21 unit cells), and for simplicity, observe light propagation before the radiated waves are reflected from the boundaries of the PC.

\begin{figure}
\includegraphics[width=8.5cm]{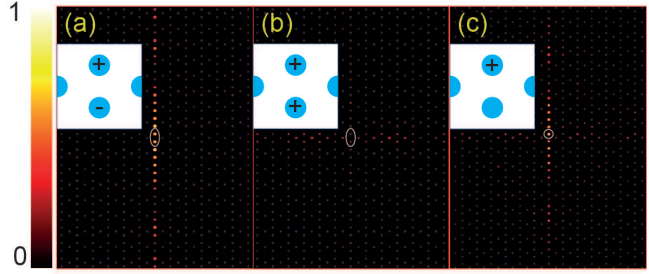}
\caption{Field intensity distributions with equal-amplitude (a) out-of-phase and (b) in-phase point sources, and with (c) a single point source. All sources are tuned to the same frequency $f=0.501107$. The location of the point sources are presented by the ellipses. Insets: Schematic diagrams for the given sources. All figures are plotted on the same scale, normalized by the maximum intensity of (a).} \label{fg:collimation}
\end{figure}

Figure \ref{fg:collimation} shows the field intensities with in-phase and out-of-phase point sources on A and B after some time steps in the FDTD simulations. As expected, for the out-of-phase point sources, the odd-parity waves in A and B nanopillars correspond to the FB shown in Fig. \ref{fg:FBdist}(a), resulting in prohibited light propagation in the $x$ direction. Therefore, we observe an interesting optical phenomenon where the radiated light from the point sources is strongly collimated into the $y$ direction, as exhibited in Fig. \ref{fg:collimation}(a). Such a phenomenon is called ``self-collimation''---light propagation is allowed in one only direction because of vanishing group velocity in the perpendicular directions.\cite{Kosaka1999,Witzens2002,Yu2003} On the other hand, for the in-phase point sources, there is no photonic mode correlated with the even-parity waves in A and B, resulting in marginal effects in terms of light propagation (see very low field intensity in Fig. \ref{fg:collimation}(b)).

As the existence of FBs has been believed to be a consequence of destructive interference, not only opposite phases but also equal amplitudes of the electromagnetic waves are therefore required. Despite this, the self-collimation effects still remain observable even with a single point source, as displayed in Fig. \ref{fg:collimation}(c). Such a robustness of the FB-induced collimation effect is understood as follows. The out-of-phase and in-phase excitations in use for Figs. \ref{fg:collimation}(a) and (b) are represented as vector $\phi_{a}= \left(1,-1\right)/\sqrt{2}$ and $\phi_{b}=\left(1,1\right)/\sqrt{2}$; the former corresponds to the FB mode whereas the latter does not. On the other hand, the single-source excitation in Fig. \ref{fg:collimation}(c) is expressed as a vector composed of the linear combination of $\phi_{a}$ and $\phi_{b}$, i.e., $\phi_{c}=A_{1}\phi_{a}+A_{2}\phi_{b}$. One can easily see that $\phi_{a}$ is always dominant, since $\phi_{b}$ barely affects light propagation in the PC. Consequently, the self-collimation effect in the FB PC is expected to be observable with arbitrary excitations in A and B.

\section{Conclusions}

We have investigated the photonic band structure and optical properties of a two-dimensional photonic crystals, which is constructed by a periodic array of quasi-1D tunable diamond lattices. It has been shown that the 2D PC exhibits flat bands only along the $\Gamma\rightarrow X$ line in the Brillouin zone, reflecting the existence of compact localized states as a consequence of destructive interference between the odd-parity waves in the A and B nanopillars. Thus, we can conclude that the FBs of the TD lattice remain even if the TD lattices are not isolated. We have also found that FBs occur not only for the fundamental modes but also for excited modes in the nanopillars. Further, we have learned that the FBs are not involved in optical transmittance through the PC because of their nondispersive nature. Moreover, by changing the direction of the incident plane wave, the anisotropy of the photonic bands and transmittance spectra were investigated. This anisotropic property of the 2D PC leads to an intriguing optical phenomenon---the self-collimation of light propagation, resulting from the existence of the nondispersive FB .

Flat band-induced strong localization acquired in disorder-free PCs can be widely employed to send photons over long distances with optical fiber communication technology. Also, our finding of a self-collimation effect at FB frequency may be practically applicable to photonic and metamaterial research. 

\begin{acknowledgement}
This work was supported by Project IBS-R024-D1, the National Research Foundation of Korea (NRF) grant funded by the Korea government (MSIT) (No. 2017076824), and research funds from Chosun University 2017.
\end{acknowledgement}

\bibliography{Q1DArrayPC}

\end{document}